\pdfoutput=1
\documentclass[aps,pre, twocolumn, groupedaddress]{revtex4-1}
\usepackage{lipsum}
\usepackage{mathtools}
\usepackage{graphicx}
\usepackage{dcolumn}
\usepackage{amsmath}    
\usepackage{amssymb}
\usepackage{bm}
\usepackage{hyperref}
\usepackage{latexsym}
\usepackage{verbatim}
\usepackage{color}
\usepackage{float}
\usepackage[caption=false]{subfig}
\usepackage{epstopdf}

\def\beq{\begin{equation}}
\def\eeq{\end{equation}}

\def\beq{\begin{equation}}
\def\eeq{\end{equation}}
\def\bea{\begin{eqnarray}}
\def\eea{\end{eqnarray}}

\DeclareRobustCommand{\uvec}[1]{{%
  \ifcsname uvec#1\endcsname
     \csname uvec#1\endcsname
   \else
    \bm{\hat{\mathbf{#1}}}%
   \fi
}}

\preprint{}

\begin{document}

\title{Polar swimmers induce several phases in active nematics}

\author{Pranay Bimal Sampat}
\email[]{pranayb.sampat.phy16@iitbhu.ac.in}
\affiliation{Department of Physics, Indian Institute of Technology (BHU), Varanasi, U.P. India - 221005}

\author{Shradha Mishra}
\email[]{smishra.phy@iitbhu.ac.in}
\affiliation{Department of Physics, Indian Institute of Technology (BHU), Varanasi, U.P. India - 221005}

\begin{abstract}
Swimming bacteria in passive nematics in the form of lyotropic liquid crystals are defined 
	as a new class of active matter known as living liquid crystals in recent studies. 
	It has also been shown that liquid crystal solutions are promising candidates for trapping 
	and detecting bacteria. We ask the question, can a similar class of matter be 
	designed for background nematics which are also active? Hence, we developed a minimal model 
	for the mixture of polar particles in active nematics. It is found that the active nematics 
	in such a mixture are highly sensitive to the presence of polar particles, and show the 
	formation of large scale higher order structures for a relatively low polar particle density. Upon increasing the density of polar particles, different phases of active nematics are found and it is observed that the system shows two phase transitions. The first phase transition is a first order transition from quasi-long ranged ordered active nematics to disordered active nematics with larger scale structures. On further increasing density of polar particles, the system transitions to a third phase, where polar particles form large, mutually aligned clusters. These clusters sweep the whole system and enforce local order in the nematics. The current study can be helpful for detecting the presence of very low densities of polar swimmers in active nematics and can be used to design and control different structures in active nematics.

\end{abstract}

\maketitle
\section{Introduction}
\label{introduction}
\begin{figure*}
\begin{tabular}{ccc}
\fbox{\includegraphics[width=5.5cm]{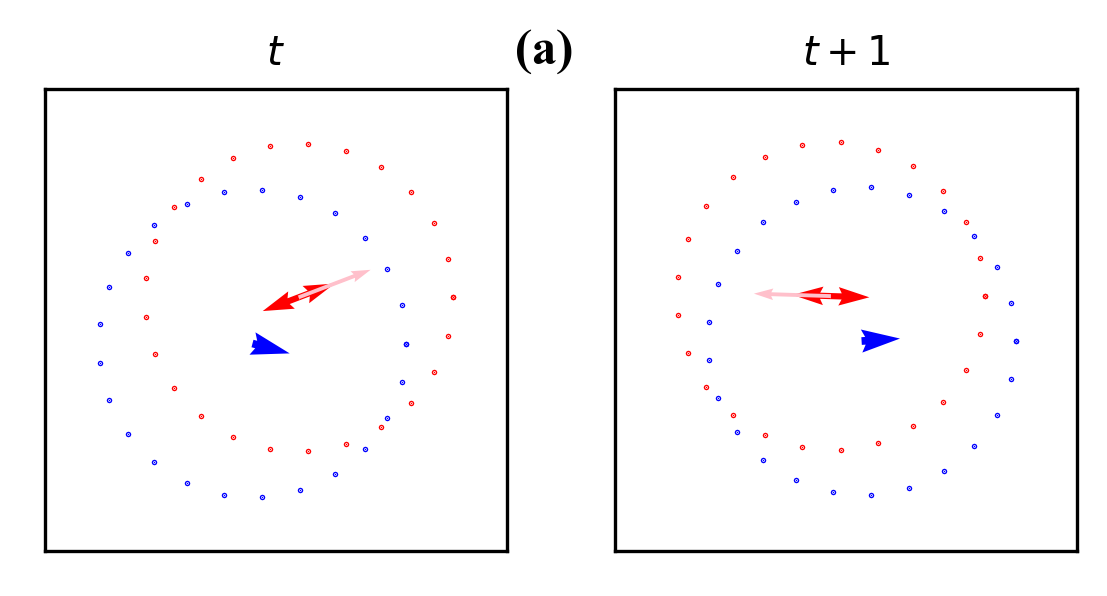}}&\fbox{\includegraphics[width=5.5cm]{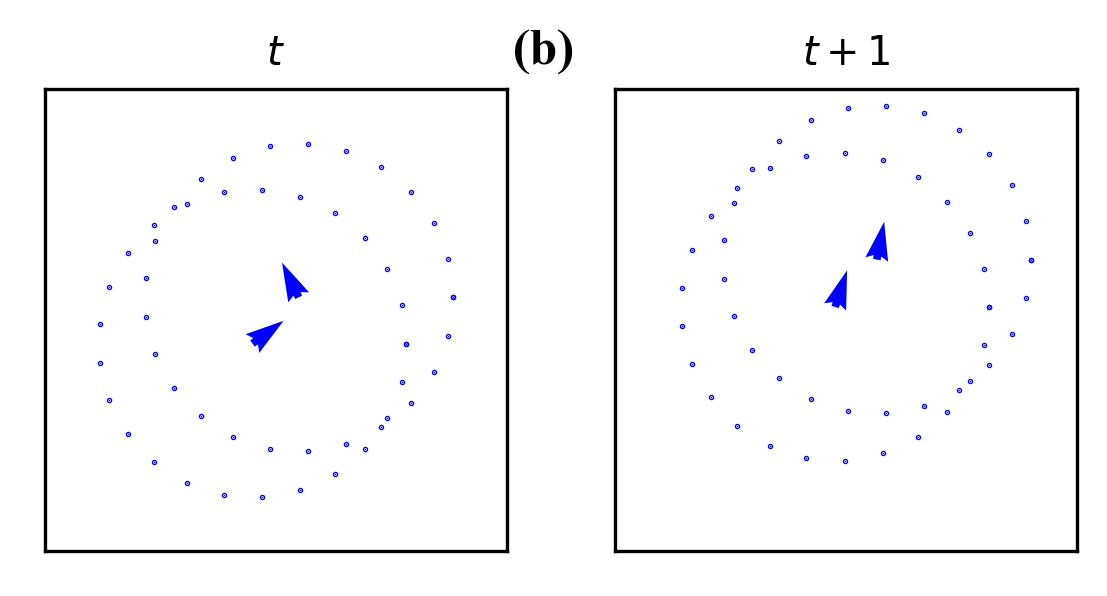}}&\fbox{\includegraphics[width=5.5cm]{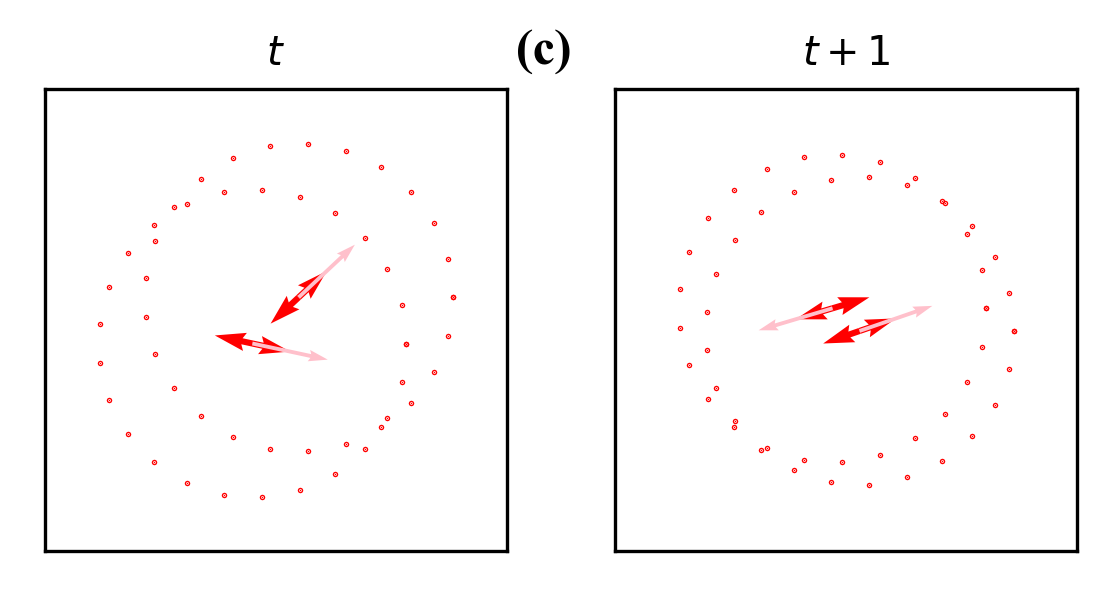}}\\
\end{tabular}
\caption{A cartoon of the model showing the different kind of interactions between particles in the mixture. Polar swimmers are represented as blue quivers, and the instantaneous direction of their head is represented through the direction in which the tip of the quiver is pointing. Apolar rods are represented by the red figures. The pink arrows represent the instantaneous velocity of the rods. These cartoons were taken from a simulation, and the dotted circle surrounding the figures represent their circles of interaction. (a) Interaction between apolar rods and polar swimmers. (b) Interactions between polar swimmers (c) Interactions between apolar rods. Although several particles may exist within the interaction range, for simplicity the cartoon shows only two particles interacting at a time.}
\label{modelfig}
\end{figure*}
Collective behaviour of self-propelled particles shows a number of interesting features, which are generally not present in the corresponding equilibrium analog \cite{sriramrev2005, rmp2013, vicsekphysicsreport2012}. Examples of such systems range from very small scales like the collection of  molecular  motors,   cytoskeletal  filaments  \cite{gruler,  goldstein}, colonies  of  swarming  bacteria  \cite{li}, much larger scales like bird flocks, fish schools etc., and many artificially designed systems like self-propelled rods \cite{vijay2007,motilecolloids}. The constituents of these systems are self-propelled or ``{\em active}'' and have their own sources of energy which they use to sustain their motion.\par
Previous studies of active matter mainly focused on pure active systems\cite{vicsekphysicsreport2012, chatepre2008, fily2012, mesoscopicnematics, vicsekmodel}. However, systems observed in nature can often be described as a mixture of multiple species. The interaction between agents of different species may lead to states that are not seen in purely homogenous systems of the constituent species \cite{rmp2016,preshambhavichiral,rmp2013, segdynamics}. Studies investigating the interaction between different types of active matter with different kinetics and alignment tendencies have shown interesting dynamics and steady state features. The introduction of a foreign species can significantly influence the dynamics and ordering in systems of active particles \cite{hagansoftmatter,colloidalflocks, chepizhko,yllanes, quint, sandor,quenchedrotators, spiralnematics, kummel2015, harder2014, kaiser2015, kaiser2014}. Such techniques have not just been explored in theoretical studies, but they have also been verified by their experimental counterparts \cite{angelani2011, activerodbeadmix, activecellbeadmix, bacterialswarmsmix}.\par

Recent studies have shown that the introduction of polar active matter in  the form of bacteria to passive liquid crystals leads to an interesting class of liquid crystal called as living liquid crystals (LLC) \cite{zhoullc1,  swimmerinteraction, commandtopologicaldefects, zhoullc2, mediumanisotropy, llchyd,llcpolarjets}. The introduction of bacteria to the liquid crystal can lead to the local melting of the liquid crystal and the formation of topological defects. LLCs are great candidates for applications such as the guided motion of bacteria and for biosensoring.\par

In LLCs the molecules forming the liquid crystal do not have any intrinsic {\it activity}. The active analog of such a liquid crystal, known as ``{\em active nematics}'' has been  observed in many  biological systems of varying scales and types \cite{activenematics1, activenematics2, activenematics4, activenematics5,activenematics6, epi1}. Motivated by the studies of LLCs and the interesting properties of the active nematics, we  study the effect of polar active particles in an active nematic medium. To do so, we introduce a minimal model of  a mixed system of active  apolar rods and polar  particles which we refer to as {\em polar swimmers}.\par

The aim of this study is to understand and enumerate how the introduction of polar swimmers to a system of active nematics affects the ordering and
structural properties of the active nematics. While the study was motivated by the work done on LLCs, we have chosen to model our polar swimmers as polar particles and not polar rods. This is because polar rods would have the same alignment mechanism as the active apolar rods in the study, and such a mixture might not affect the nematic ordering in the system. We tune the density of polar swimmers in the background of a dense system of apolar rods.
The primary result of this study is that a very small density of polar swimmers is enough to break the quasi-long ranged ordered state
seen in pure active nematics \cite{chateminimal}. A non-monotonic change in apolar ordering, with the density of polar swimmers is observed. Such a mixture can broadly be divided into three phases, depending on the concentration of the polar swimmers. Interestingly, higher order structures are observed in the mixture, which are generally not present in pure active nematics \cite{nematicdefect1, nematicdefect2, nematicdefect3}.\par

The rest of the article is divided in the following manner: In section \ref{section:model} we give the details of the model and the numerical simulation. In section \ref{section:results} we present the results of our study, and finally in section \ref{section:discussion} we summarise our results and discuss future directions of the problem.

\section{Model and Numerical Details}
\label{section:model}

Our system consists of a mixture of active apolar rods and polar particles  moving on a two-dimensional substrate of friction coefficient $\mu$. Each particle is driven by an internal force $F$ acting along the long axis of the particle. The ratio of the force and the friction coefficient gives a constant self-propulsion speed $v_0=\frac{F}{\mu}$. Each particle (apolar/polar) is a {\em point particle} defined by a position vector ${\bf r}_{a/p}(t)$ and orientation $\theta_{a/p}(t)$. The motion of apolar rods is symmetric with respect to the angle $\theta_a = \theta_a + \pi$, whereas the polar particles in the model  move assymetrically along the direction of their orientation $\theta_p$. Both types of particles align with their neighbours within a  radial distance from them, less than a fixed radius of interaction $r_0$. The polar particles, coined as polar swimmers in this study, are defined such that they align ferromagnetically with the direction of velocity of polar particles and with the instantaneous direction of velocity of apolar rods. The alignment interaction is similar to the minimal model suggested by Vicsek et al in \cite{vicsekmodel}. Apolar rods align their orientation in an apolar fashion with the polar swimmers and apolar rods in their neighbourhood \cite{chateminimal, mesoscopicnematics}. The cartoon of the three types of interactions: (i) apolar-polar, (ii) polar-polar and (iii) apolar-apolar is shown in Fig. \ref{modelfig}(a-c), respectively.\par

The system is studied on a  $L \times L$ square geometry and the periodic boundary condition is used in both directions.
The orientation and the position of the apolar rods are updated at a unit time interval as follows:

\begin{equation}
\vec{v}_{i,a}(t)=R_{i}(t)v_{0}(\cos(\theta_{i,a}(t))\hat{x}+\sin(\theta_{i,a}(t))\hat{y}) \ ,
\label{apolvel}
\end{equation}
\begin{equation}
\vec{r}_{i,a}(t+1)=\vec{r}_{i,a}(t)+\vec{v}_{i,a}(t) \ ,
\label{apolposupd}
\end{equation}
\begin{equation}
\theta_{i, a}(t+1)=\frac{1}{2}\arg\left(\left<e^{i2\theta (t)}\right>_{\vec{r}_{i,a}(t),r_{0}}\right)+\eta_{i,t} \ .
\label{apolvelupd}
\end{equation}
$R_{i}(t)$ is randomly chosen for each particle at each time-step to be $-1$ or $+1$ with equal probablity (apolar rods can move with equal probability along $\theta$ and $\theta+\pi$).
The argument of the term $\frac{1}{2}\left<e^{i2\theta (t)}\right>_{\vec{r}_{i,a}(t),r_{0}}$ is a measure of the average {\it orientation} of both apolar rods and polar swimmers within a
circle centered at $\vec{r}_{i,a}(t)$ and within the neighbourhood  radius $r_0$.
\begin{equation}
\begin{aligned}
\left<e^{i2\theta (t)}\right>_{\vec{r}_{i,a}(t),r_{0}}=\sum\limits_{\lvert \vec{r}_{j,a}-\vec{r}_{i,a} \rvert <r_{0}}{e^{i(2\theta_{j,a}(t))}} \\ +\sum\limits_{\lvert\vec{r}_{j,p}-\vec{r}_{i,a}\rvert<r_{0}}{e^{i(2\theta_{j,p}(t))}}
\end{aligned}
\label{avgorient}
\end{equation}

The orientation and position of  the polar swimmers are updated as follows:

\begin{equation}
\vec{v}_{i,p}(t)=v_{0}(\cos(\theta_{i,p}(t))\hat{x}+\sin(\theta_{i,p}(t))\hat{y}) \ ,
\label{polarvel}
\end{equation}
\begin{equation}
\vec{r}_{i,p}(t+1)=\vec{r}_{i,p}(t)+\vec{v}_{i,p}(t) \ ,
\label{polarposupd}
\end{equation}
\begin{equation}
\theta_{i, p}(t+1)=\arg\left(\left<\vec{v}(t)\right>_{\vec{r}_{i,p}(t),r_{0}}\right)+\eta_{i,t} \ .
\label{polarvelupd}
\end{equation}
The argument of $\left<\vec{v}(t)\right>_{\vec{r}_{i,p}(t),r_{0}}$ is a measure of the average instantaneous velocity of all particles
within a circle of interaction centered at $\vec{r}_{i,p}(t)$ and within a neighbourhood  radius $r_0$.
\begin{equation}
\left<\vec{v}(t)\right>_{\vec{r}_{i,p}(t),r_{0}}=\sum\limits_{\lvert\vec{r}_{j,a}-\vec{r}_{i,p}\rvert<r_{0}}{{\vec{v}_{j,a}(t)}} +\sum\limits_{\lvert\vec{r}_{j,p}-\vec{r}_{i,p}\rvert<r_{0}}{\vec{v}_{j,p}(t)}
\label{avgvel}
\end{equation}

The alignment interactions introduced in eqs. \ref{apolvelupd}  and \ref{polarvelupd} are purely motivated by the instantaneous collision of two particles moving along their long axis. The term $\eta$ refers to a delta-correlated noise distributed uniformly over the range [-$\eta/2, \eta/2$]. This noise is representative of the random error
in the alignment of the particles. The magnitude of velocity (speed) of both the species was fixed as $v_0=1/4$. The density of nematic rods ($\frac{N_{a}}{L^2}$)
fixed as $1.0$. The density of polar swimmers ($\frac{N_{p}}{L^2}$), denoted by $\rho_p$, is the primary variable of the study, and varied in the range [$0,0.5$] ($L$ is kept fixed and  and $N_p$ is varied). The noise strength $\eta$ used is in the range $[0.1, 0.3]$, which is in the ordered phase of a pure active nematic system with the chosen density.\par

We start with a random homogeneous distribution of apolar rods and polar swimmers, and their position and orientation are updated as per eqs. \ref{apolvel}-\ref{avgvel}.
One simulation time-step is counted as an update of all $N = N_a+N_p$ particles in parallel. At least $8$ realisations of each configuration of $(L, \eta, \rho_p)$
were simulated for better averaging. Each realisation was simulated for $2 \times 10^5$ timesteps, of which the first $1.5 \times 10^5$ were used to
anneal the system to the corresponding steady state, and the data from the remaining time-steps were used to calculate the results in the study.
In the critical region, at low densities of $\rho_p$, $10$ additional realisations of the system were simulated for $4 \times 10^5$ time-steps,
of which the first $1.5 \times 10^5$ were used to anneal the system to the corresponding steady state.\par

\section{Results}
\label{section:results}

\subsection{Nematic Ordering}

\begin{figure}[ht]
\includegraphics[height=8.5cm]{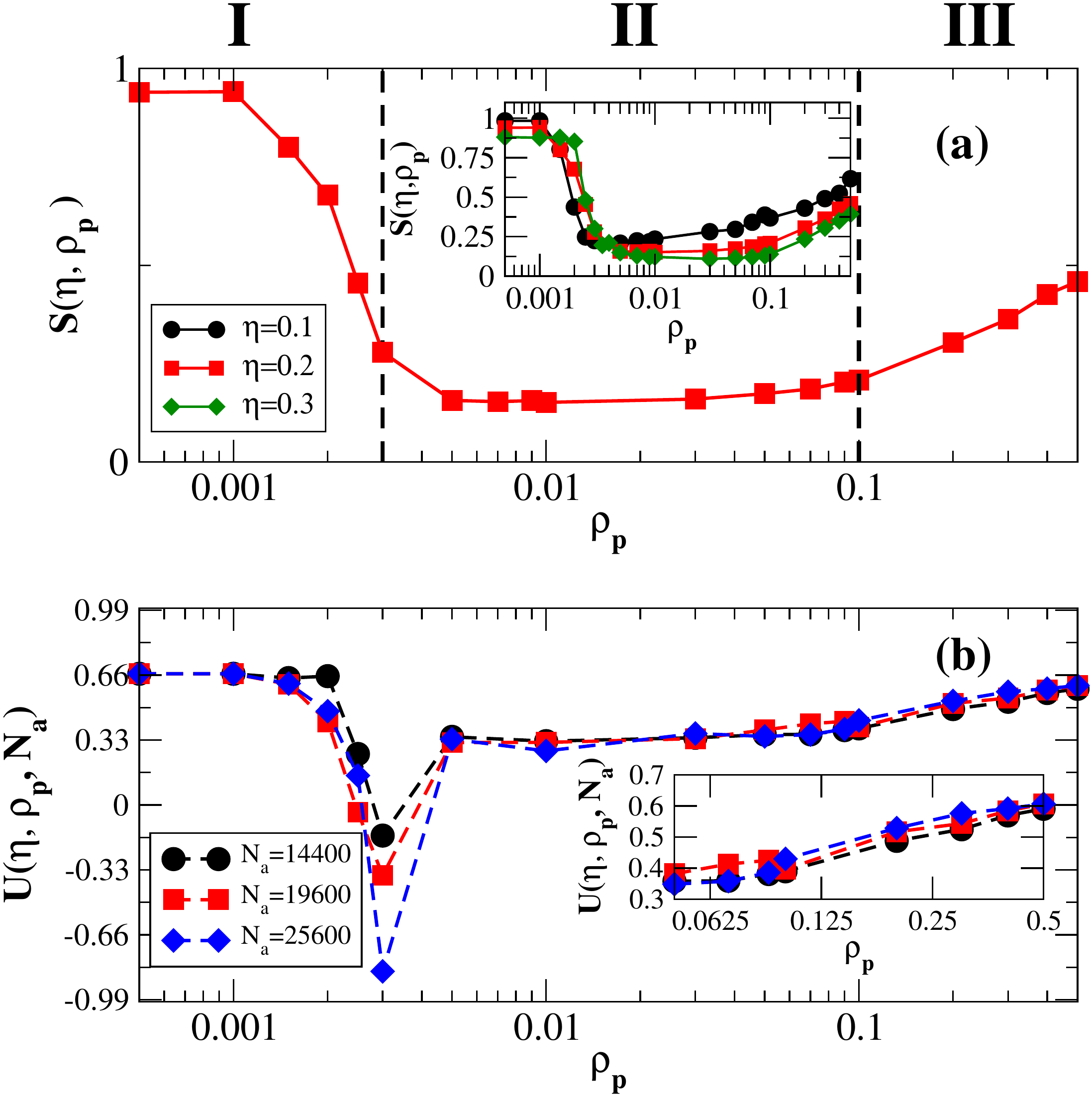}
\caption{(a) Variation of $S(\eta, \rho_p)$ with $\rho_p$ for $\eta=0.2$ and $L=160$. Three clear regions are observed and marked as ($I, II$ and $III$). The two dashed lines show the division between the three distinct regions. (Inset) The same non-monotonic behaviour is observed for all values of $\eta = 0.1,0.2$ and $0.3$. (b) The variation of Binder cumulant $U(\eta,\rho_p)$ with $\rho_p$ for $\eta=0.2$. The upper and lower  dashed lines corresponds to the  $U(\eta, \rho_p)=\frac{2}{3}$ and $\frac{1}{3}$ for the ordered and disordered state respectively. The inset in (b) shows the zoomed in plot of $U(\eta, \rho_p)$ near the second phase transition.}
\label{scalarnematic}
\end{figure}

We first calculate the variation in the global nematic ordering with the density of polar swimmers. The ordering in the system is quantified by the scalar nematic order parameter $S(\eta, \rho_p)$,  defined as:
\begin{equation}
S(\eta, \rho_p) = \left<\lvert \frac{1}{N_a} \sum\limits_{j=1}^{N_{a}}{e^{i2\theta_{j,a}}} \rvert_{\eta, \rho_p}\right>_{t}
\label{scalarnematicdefinition}
\end{equation}

where $\left<...\right>_{t}$ implies averaging over multiple realisations and a number of time steps in the steady state. The value of $S(\eta, \rho_p)$ is $\simeq 1$ for an ordered state and is $\sim 0$ for a disordered state. In Fig. \ref{scalarnematic} we plot $S(\eta, \rho_p)$ vs. $\rho_p$, for three different noise strengths $\eta=0.1, 0.2$ and $0.3$. The plot shows the non-monotonic variation of $S(\eta, \rho_p)$ with $\rho_p$ and three distinct regions. These different regions are actually the three  distinct phases of the mixture, which will be discussed ahead.

\begin{figure}[H]
\begin{tabular}{c}
\fbox{\includegraphics[width=7.75cm]{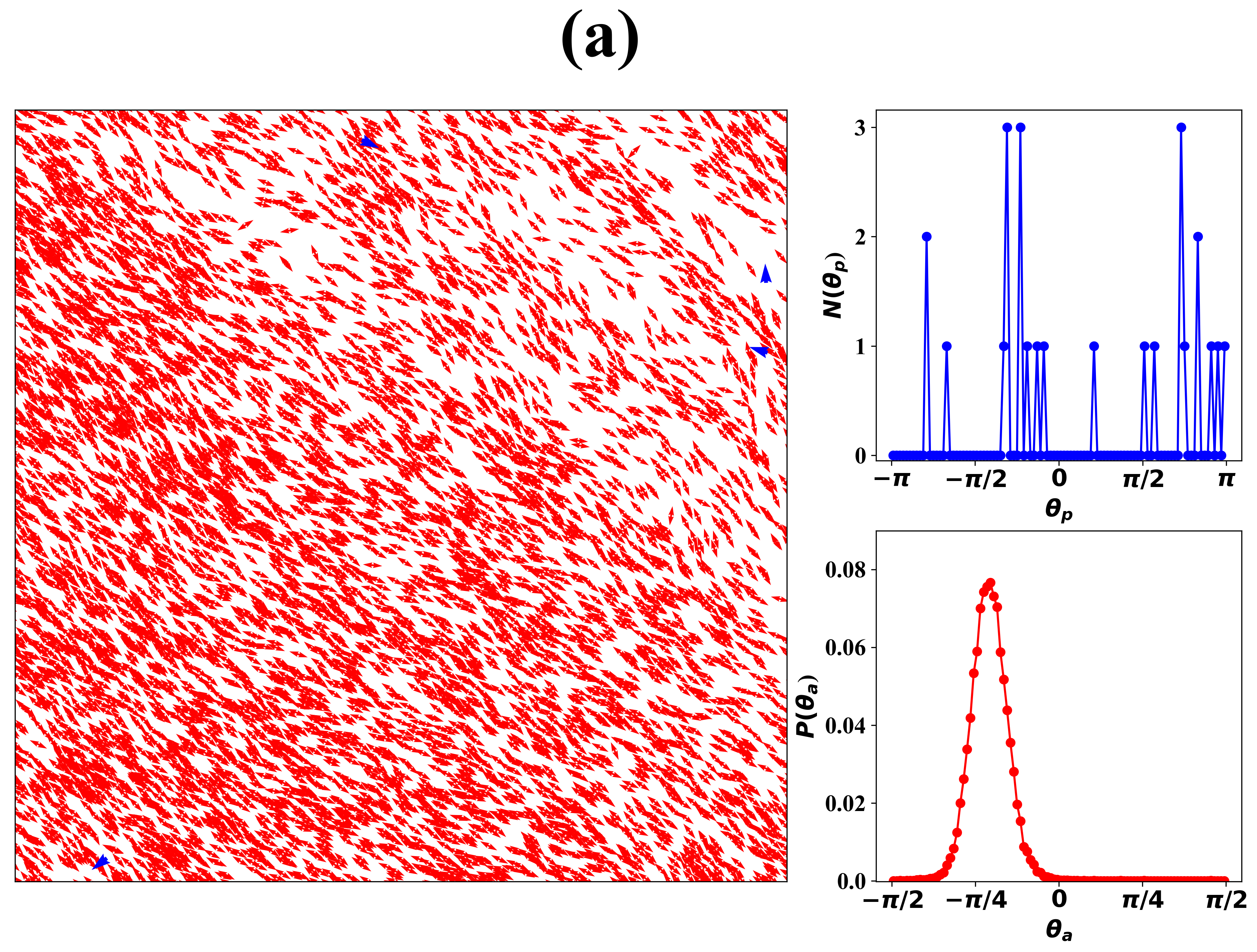}}\\\fbox{\includegraphics[width=7.75cm]{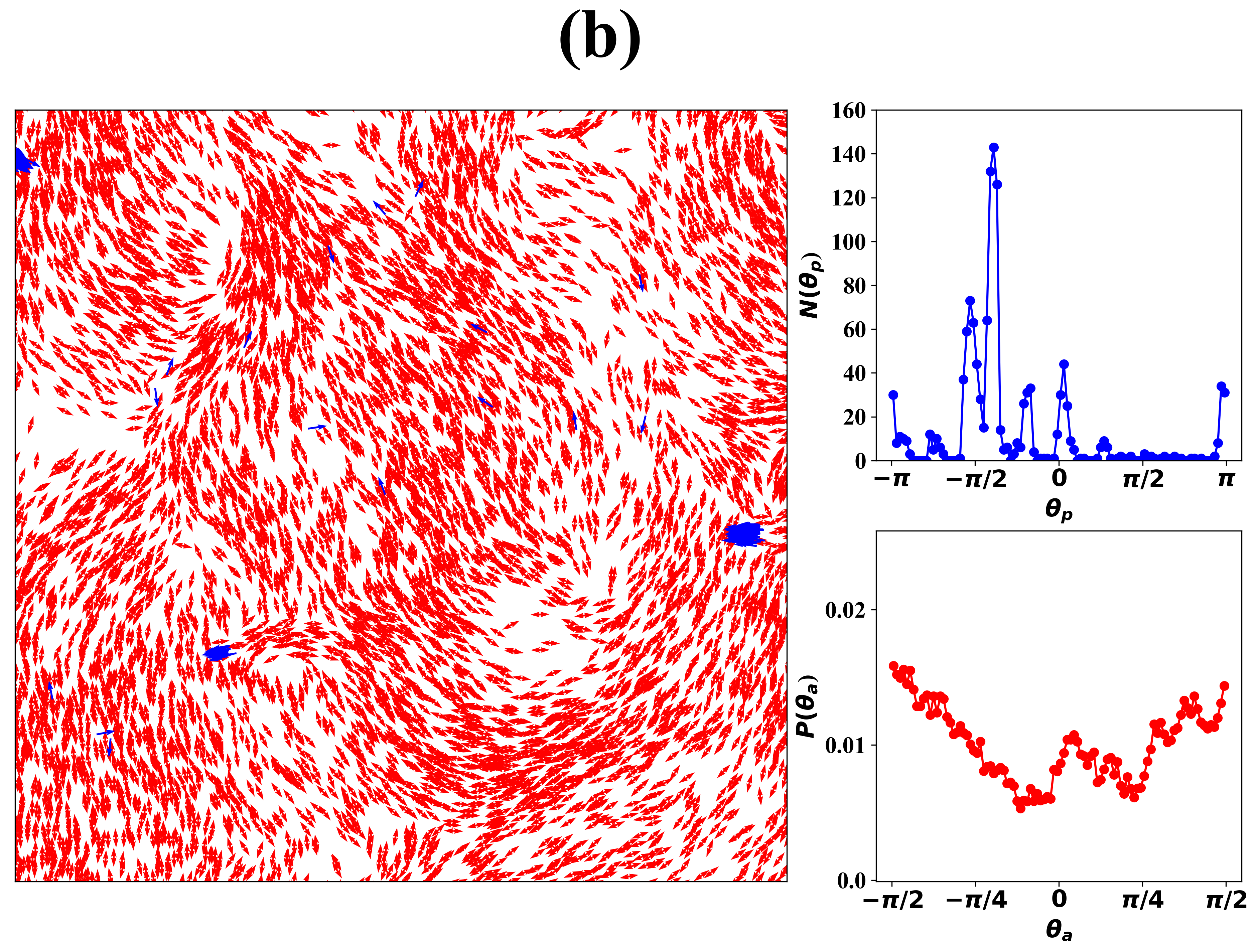}}\\\fbox{\includegraphics[width=7.75cm]{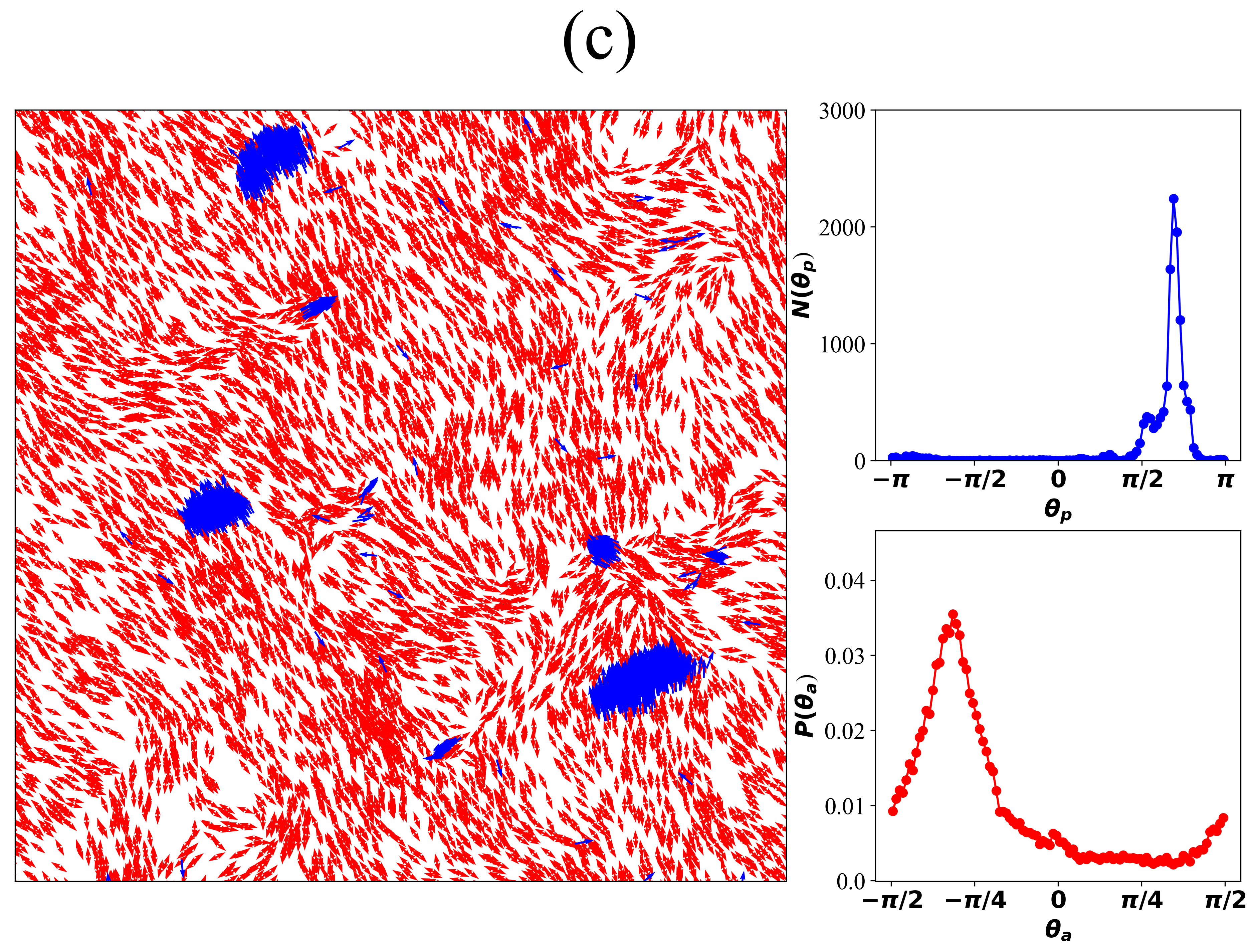}}\\
\end{tabular}
\caption{Real space snapshots of apolar rods (red) and polar swimmers (blue) for $L=160$ in the (a) first phase ($\rho_p=0.001$) (b) second phase ($\rho_p=0.05$) and (c) third phase ($\rho_p=0.5$). The snapshots show one quarter of the whole system. The panel on the right in each figure shows the number distribution (ND) of polar swimmers $N(\theta_p)$ (top) and probability distribution function (PDF) of apolar rods $P(\theta_a)$(bottom) for the corresponding snapshot.}
\label{snapshots}
\end{figure}

The  snapshots of apolar rods (red) and polar swimmers (blue) are also shown in Fig. \ref{snapshots}. The distribution of orientation of the rods is shown by plotting the probability distribution function (PDF) $P({\theta_a})$ and the number distribution (ND) of polar swimmers by  $N(\theta_{p})$.  The $y-axis$ shows the  number of polar swimmers participating in each peak of $N(\theta_p)$. It gives a measure of the size of the clusters of polar swimmers.\par
The first phase $(I)$ is observed at {\it very} low values of $\rho_p$, and exhibits strong nematic ordering. The value of the nematic order parameter
$S$ does not deviate from the value observed in a system consisting only of apolar rods with the same density. Snapshots of the system in the steady state in this phase also show that the system forms dense bands of apolar rods, which evolve over large timescales. These features are similar to those seen in a system consisting purely of apolar rods, as confirmed by Chaté et al in \cite{chateminimal}.The PDF $P(\theta_a)$ in this phase also shows a normal distribution around a sharp peak, which is an indication of strong global ordering of the nematic rods. The peak in the distribution shows the direction of nematic ordering in the system. This suggests that the collective properties of apolar rods are not affected by the presence of a  small density of polar swimmers.
In this phase polar swimmers do not interact with other swimmers and hence do not form clusters. The ND $N(\theta_p)$ shows random spikes at different values of $\theta_p$ as shown in Fig. \ref{snapshots}.(a)\par

\begin{figure}
\centering
\begin{tabular}{c}
\fbox{\includegraphics[width=7.7cm]{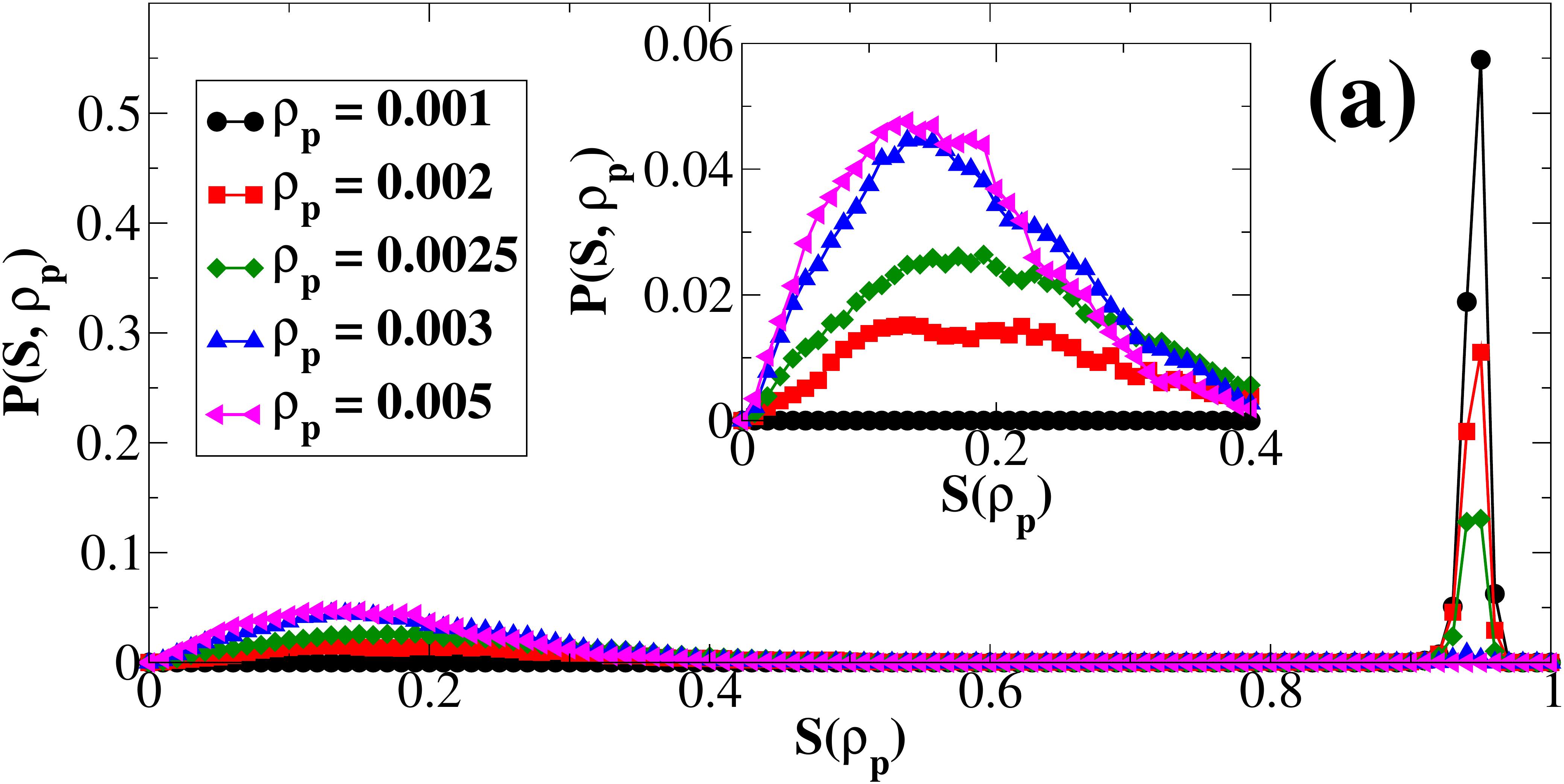}}\\
	\fbox{\includegraphics[width=7.7cm]{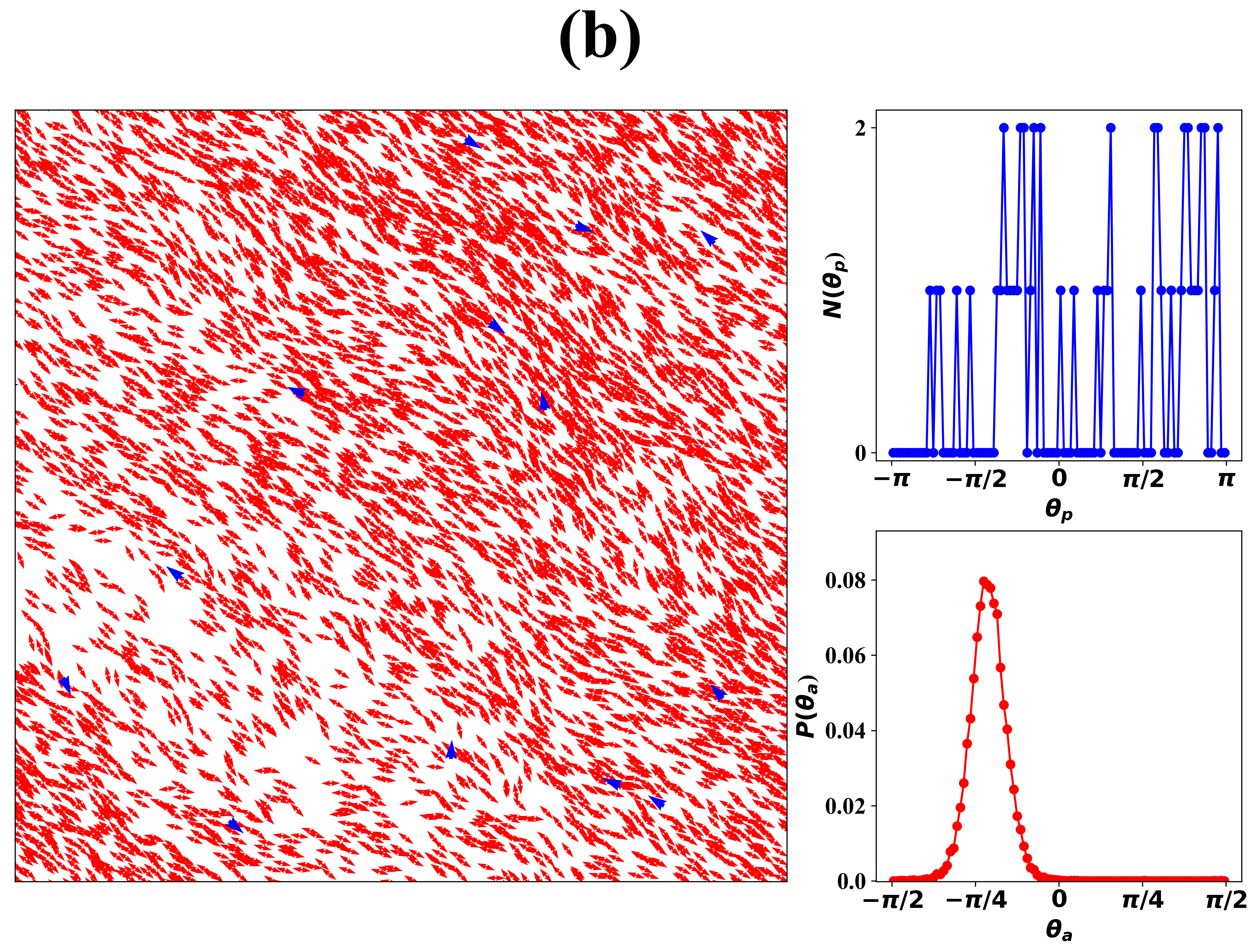}}\\
\fbox{\includegraphics[width=7.7cm]{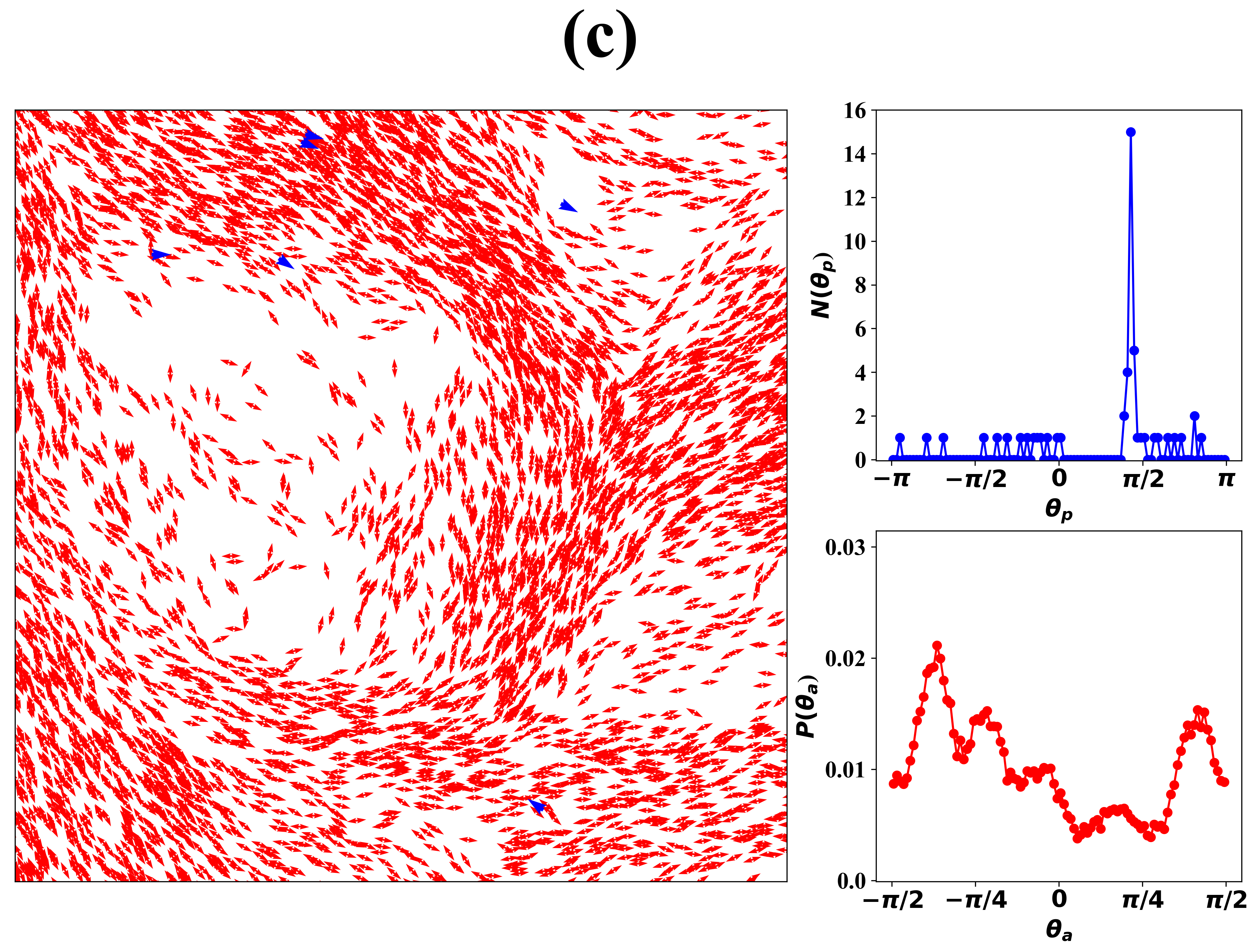}}
\end{tabular}
\caption{(a) The steady state probability distribution function $P(S)$ of $S$ in the steady state for different values of $\rho_p=0.001, 0.002, 0.0025,0.003$ and $0.005$	near the critical point of first and second phase for $\eta=0.2$. There is a a clear bimodal distribution for $\rho_p=0.002$, $\rho_p=0.0025$ and $\rho_p=0.003$. The inset plot shows the systematic emergence of a second peak in the range of $S$ for disordered nematics. (b) A snapshot of a nematically ordered realisation for $L=160$, $\rho_p=0.002$, and $\eta=0.2$  and (c) a snapshot of a nematically disordered realisation for the same conditions. Both snapshots show one quarter of the whole system. The panel on the right in figure (b) and (c) shows the number distribution (ND) of polar swimmers $N(\theta_p)$ (top) and probability distribution function (PDF) of apolar rods $P(\theta_a)$(bottom) for the corresponding snapshot.}
\label{bimodal}
\end{figure}

A sharp drop in $S$ is observed as $\rho_p$ is increased beyond a threshold value. This drop is followed by a flat region where $S$ is invariant to further increase in $\rho_p$. This flat region in the plot for $S$ vs $\rho_p$ is described as the second phase of the mixture $(II)$. The transition from the first phase to the second phase is identified as a first order transition through the Binder cumulant $U(\eta,\rho_p)$ and the probability distribution function (PDF) of scalar order parameter $P(S)$ in Fig. \ref{bimodal}(a). $U(\eta, \rho_p)$ is defined as $U=1-\frac{\left<S^4\right>}{3\left<S^2\right>^2}$. Fig. \ref{scalarnematic}(b) shows the variation of $U(\eta,\rho_p)$  with $\rho_p$ for noise strength $\eta=0.2$. $U(\eta,\rho_p)$ shows a sharp drop to negative values near the transition from phase $I$ to $II$ with the magnitude of the drop increasing with an increase in system size, which is characteristic of a first order transition. The bimodal distribution of $P(S)$ for $\rho_p$ in the critical range of $\rho_p = 0.002 $ to $0.003$ confirms the first order transition. The two peaks of bimodal distributions correspond to two distinct states: ordered and disordered, which coexist for the same system conditions. Whereas in the first ($\rho_p=0.001$) and second ($\rho_p=0.005$) phase, $P(S)$ shows one peak at $S$ close to $1$ and $0.1$ respectively. The snapshots of the two coexisting states in the critical range of the transition at $\rho_p=0.002$ is shown in Fig. \ref{bimodal}(b) and (c), where the ordered state is similar to the first phase with a distinct peak in $P(\theta_a)$, uniformly ordered nematic rods with bands and random spikes in $N(\theta_p)$. In the disordered state, the polar swimmers form a cluster consisting of a few polar swimmers, with one major peak in $N(\theta_p)$. These clusters break the background nematic ordering and lead to the formation of defects and a broad distribution of $P(\theta_a)$.
Similarly, steady state snaphots of the system in the second phase in Fig.\ref{snapshots}(b) show that the system is disordered, with the presence of  defects in nematic ordering. The size of the defects in this phase are larger than the size of the clusters of polar swimmers. The PDF $P(\theta_a)$ also shows a uniform distribution with no clear peak and multiple distinct peaks in $N(\theta_p)$.\par

As the density of polar swimmers is further increased, a third phase $(III)$ is observed (Fig. \ref{scalarnematic}). In this phase, the nematic order parameter increases with $\rho_p$. $U(\eta,\rho_p)$ shows a smooth variation from $\frac{1}{3}$ in the second phase to $\frac{2}{3}$ as system goes to the third phase (Fig. \ref{scalarnematic}(b) (inset)), which suggests a continuous transition. The polar swimmers start to form clusters with a global mean orientation. The number distribution $N(\theta_p)$ shows a distinct peak at the global mean orientation of polar swimmers. These clusters sweep the whole system repeatedly and also enforce a director for nematic ordering. However, there are always small stray clusters of polar swimmers with random orientations which hamper the macroscopic ordering  of nematic rods. The  snapshots of the system in this phase show the presence of weak nematic ordering, with much smaller defects than those seen in the second phase of the mixture. The PDF $P(\theta_a)$ also shows the emergence of a  global director of nematic ordering, but not one as strong as the one seen in the first phase.\par
\begin{figure}[h]

\begin{tabular}{c}
	\includegraphics[width=8cm]{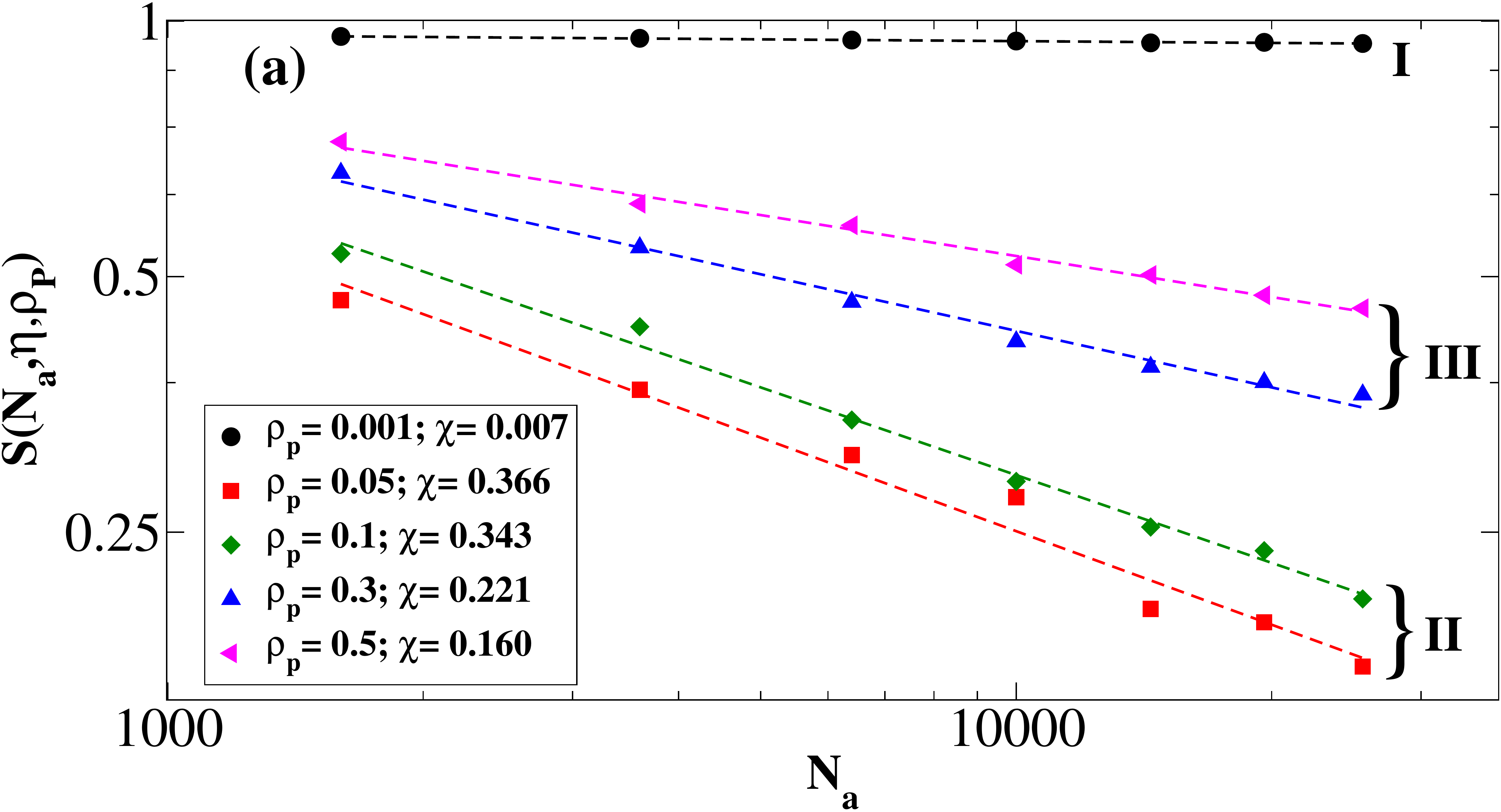}\\
	\includegraphics[width=8cm]{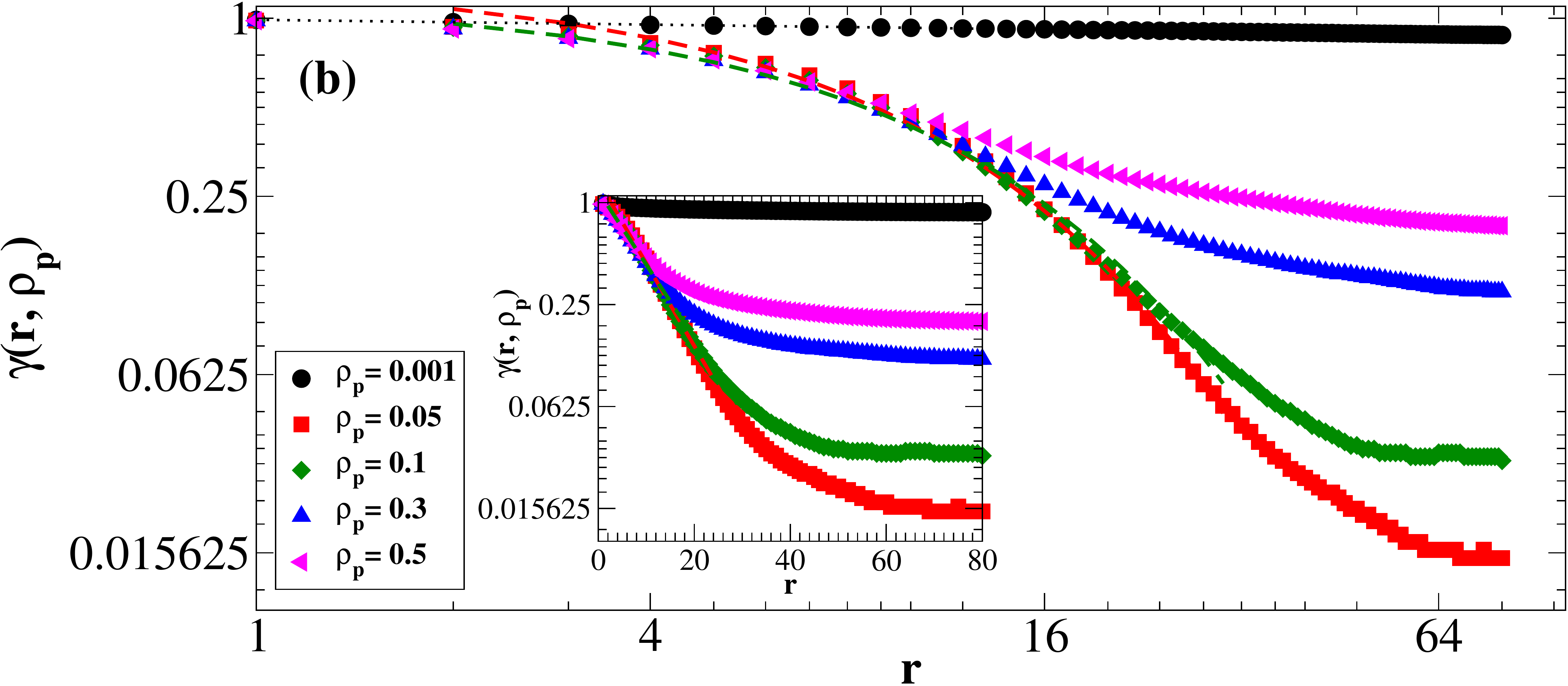}
\end{tabular}

	\caption{(a) $S(N_a,\rho_p)$ vs $N_a$ for $\eta=0.2$. $N_a$ is varied by increasing $L$, and $\rho_a$ is kept fixed as 1. The dashed line in the figure represents the fitted form $S(N_a)\sim N_a^{-\chi}$. The value of $\chi$ obtained through the fitting for the given parameters is mentioned in the legend. (b) $\gamma(r, \rho_p)$ for different values of $\rho_p$ for $N_a=25600$ and $\eta=0.2$. The dotted line represents the power law fitting $\gamma(r) \sim r^{-c_0(\rho_p)}$ and the dashed line represents the fitting $e^{-c'(\rho_p) r}$. (inset) $\gamma(r)$ vs $r$ on semilog $y-$ scale. Straight line feature of $\gamma(r)$ for distance $r \le 40$ shows the  exponential decay of correlation to  a very low value for system in the second phase.}
\label{figsn}
\end{figure}

Thus, we find that active nematic rods exist in three distinct phases in the presence of polar swimmers. To characterise the range of order in these phases we calculated the variation of $S$ with $N_a$ and the two point spatial correlation for nematic rods $\gamma(r)$, calculated as $\gamma(r)=\left< cos(\theta(r_0)-\theta(r_0+r)) \right>$, in the first, second and third phases ($\rho_p=0.001, 0.05, 0.1, 0.3$ and $0.5$). $\left<..\right>$ implies averaging over all the nematic rods $(r_0)$ across multiple realisations at multiple times in the steady-state. Fitting $S(N_a)$ to the form $S(N_a) \sim N_a^{-\chi}$, we see that the first phase exhibits quasi-long-range order in its orientation, with $\chi<\frac{1}{16}$ and a power law decay of $\gamma(r)$ \cite{chateminimal}. In the second phase, the apolar rods are disordered with $\chi \sim \frac{1}{3}$ and $\gamma(r)$ decays exponentially. The third phase exhibits short-range order with $\chi>\frac{1}{16}$, with a distinctly slower power-law decay of $\gamma(r)$ at larger distances when compared to the second phase. Hence on increasing $\rho_p$, the apolar rods undergo a non-monotonic variation of ordering from a QLRO state to a disordered state and then to a short-range ordered state. The plot of $S(N_a)$ vs. $N_a$  in the first, second,  and the third phase is shown in Fig. \ref{figsn}.

To better understand the effect of polar swimmers on the active apolar rods, we calculate the characteristics of polar swimmers in the bulk nematic.

\subsection{Characteristics of Polar Swimmers}

The characteristics of the polar swimmers in the mixture were quantified using the following parameters:

\begin{figure}[h]
\includegraphics[width=9cm]{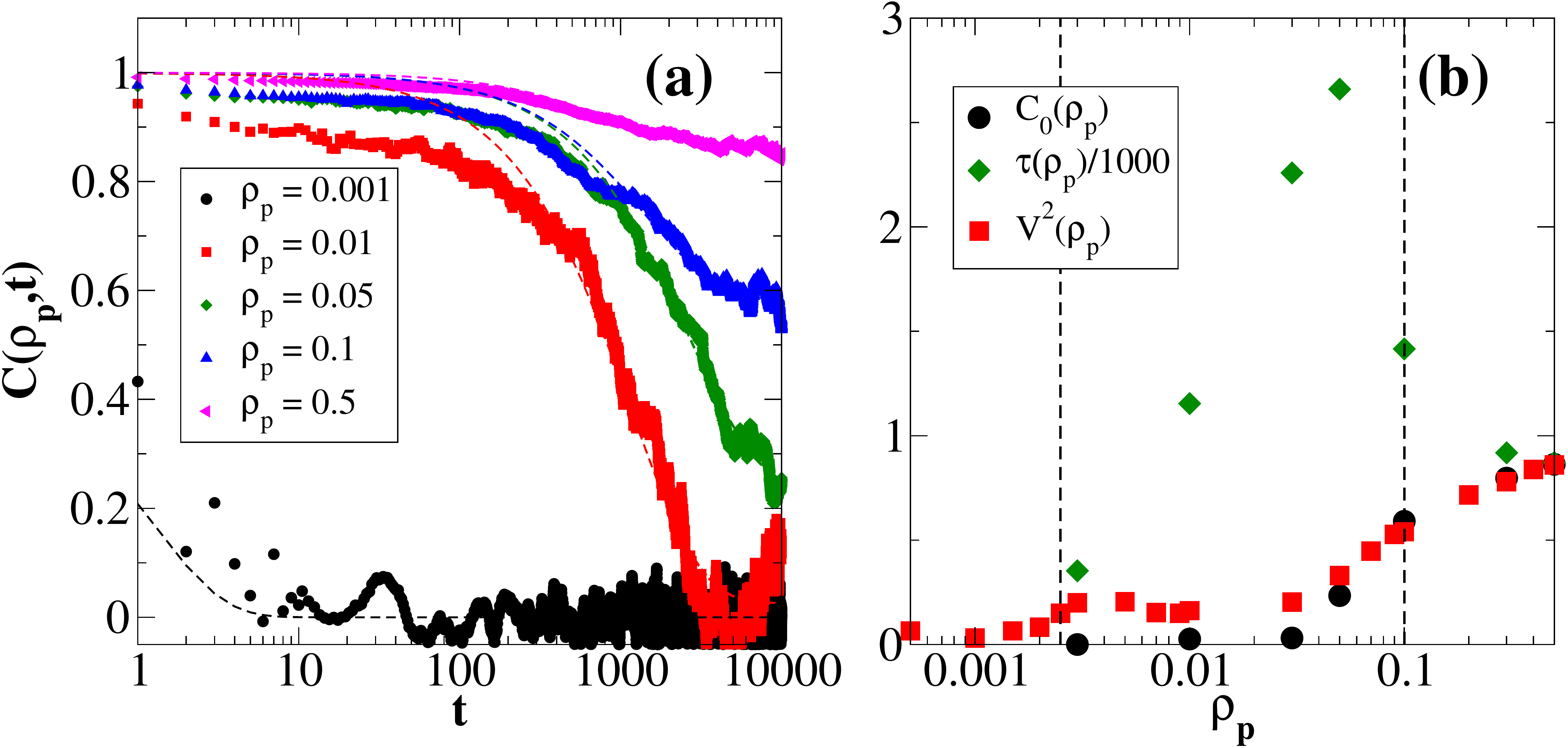}\\
\caption{(a) Polar velocity time auto-correlation function (VACF)  $C(\rho_p,t)$ for different densities of polar swimmers $\rho_p$. (b) Variation of $C_0(\rho_p)$, $\tau(\rho_p)$, and $V^2(\rho_p)$ with $\rho_p$. The two dashed lines are the boundaries between the three distinct phases, similar to Fig. \ref{scalarnematic}}
\label{vcf}
\end{figure}

1) The polar order parameter for polar swimmers $V(\rho_p)$, given as:
\begin{equation}
	V(\rho_p)=\frac{1}{v_0 N_{p}} \left< \lvert \sum\limits_{j=1}^{N_{p}}{\vec{v}_{j,p}(t)} \rvert \right>_{t}
\end{equation}
$\left<...\right>_t$ has the same meaning as given in Eq. \ref{scalarnematicdefinition}. The polar order parameter is a measure of the alignment of polar swimmers in the system. In an ordered system of polar swimmers,
where a majority of the swimmers are moving in the same direction, the value of $V(\rho_p)$ is $\simeq 1$. Alternatively, in a poorly aligned system of polar swimmers, where the swimmers move randomly, the value of $V(\rho_p)$ is $\simeq 0$.\par

2) Velocity auto-correlation function (VACF) $C(\rho_p, t)$ of polar swimmers, given as:
\begin{equation}
C(\rho_p, t)=\frac{1}{v^2_0 N_{p}} \left< \sum\limits_{j=1}^{N_{p}}{\vec{v}_{j,p}(t_0) \cdot \vec{v}_{j,p}(t_0+t)} \right> _{t_0}
\label{vacf}
\end{equation}
where $t_0$ is high enough that the system is in the steady state,  $\left<...\right>_{t_0}$ implies averaging over multiple realizations, and  $t_0$. $C(\rho_p, t)$  gives a measure of how long a swimmer remembers its orientation.\par

3) The mean-square displacement of the polar swimmers $\Delta_p(\rho_p, t)$:
\begin{equation}
\Delta_p(\rho_p,t)=\frac{1}{N_{p}}\left<\sum\limits_{j=1}^{N_{p}}{\lvert\vec{r}_{j,p}(t_0+t)-\vec{r}_{j,p}(t_0)\rvert^{2}}\right>_{t_0}
\end{equation}
where $\left<...\right>_{t_0}$ has the same meaning as in Eq. \ref{vacf}. When fitted to the form $\Delta_p(t) \sim t^{\beta(t)}$, the exponent $\beta(t) = \log_2\frac{(\Delta_p(2 t))}{\Delta_p(t)}$ can be used to identify the type of motion that the swimmers are undergoing.
$\beta(t)=1$ for diffusive motion,  $1<\beta(t)<2$ for super-diffusive motion, and $\beta(t)=2$ for ballistic motion. It should be noted that the particles undergoing ballistic motion reach the boundary of the box every $\sim 640$ steps, and hence move in the same environment as before.\par
\begin{figure}[H]
\includegraphics[width=9cm]{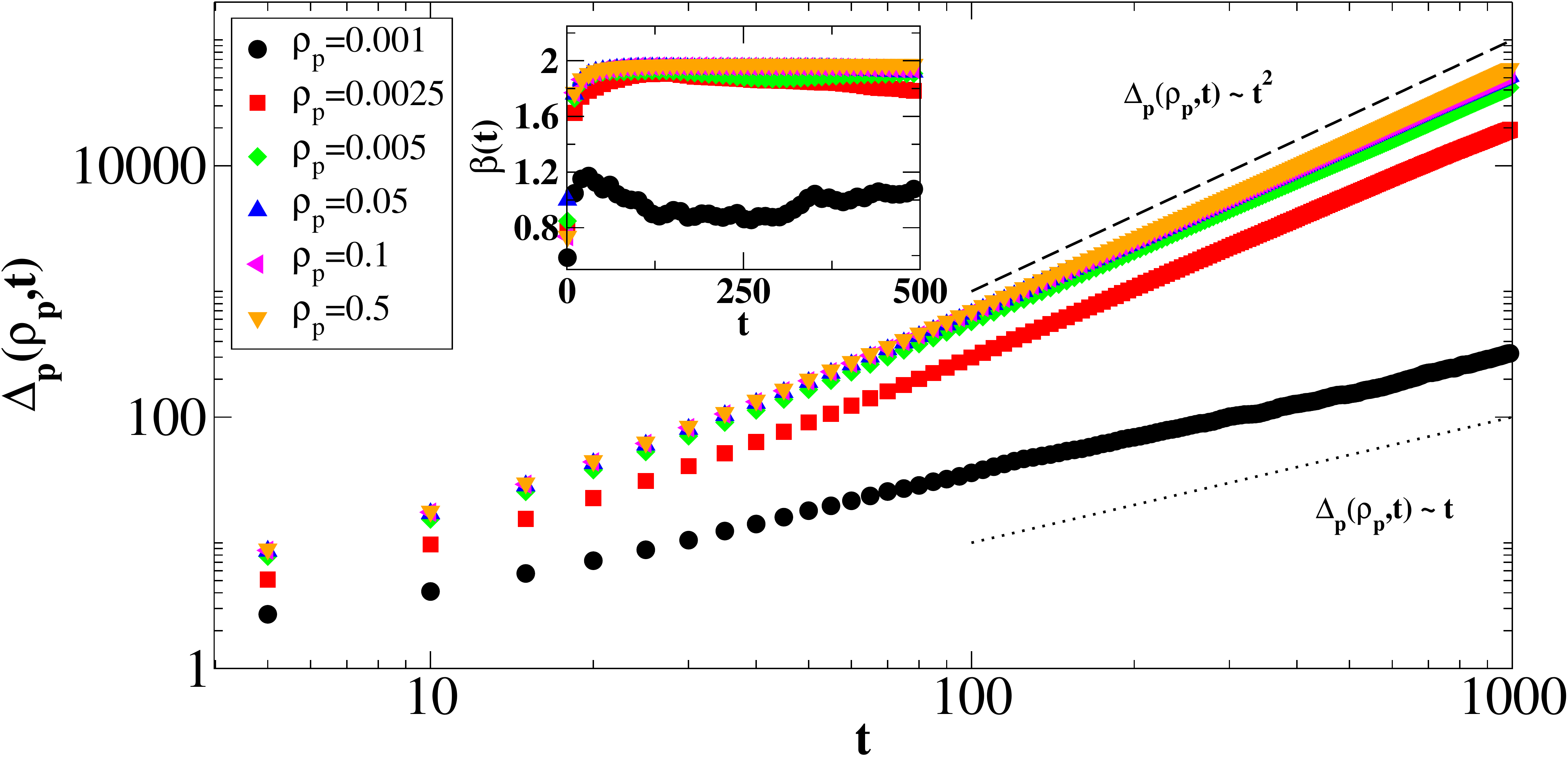}
	\caption{Variation of mean square displacement (MSD) $\Delta (\rho_p, t)$ with $t$ for different values of $\rho_p$ at $\eta=0.2$ and $L=160$. (Inset) Variation of $\beta(t)$ with $\rho_p$.}
\label{polarmsd}
\end{figure}

\begin{figure*}
\centering
\begin{tabular}{ccc}
{\includegraphics[width=6cm]{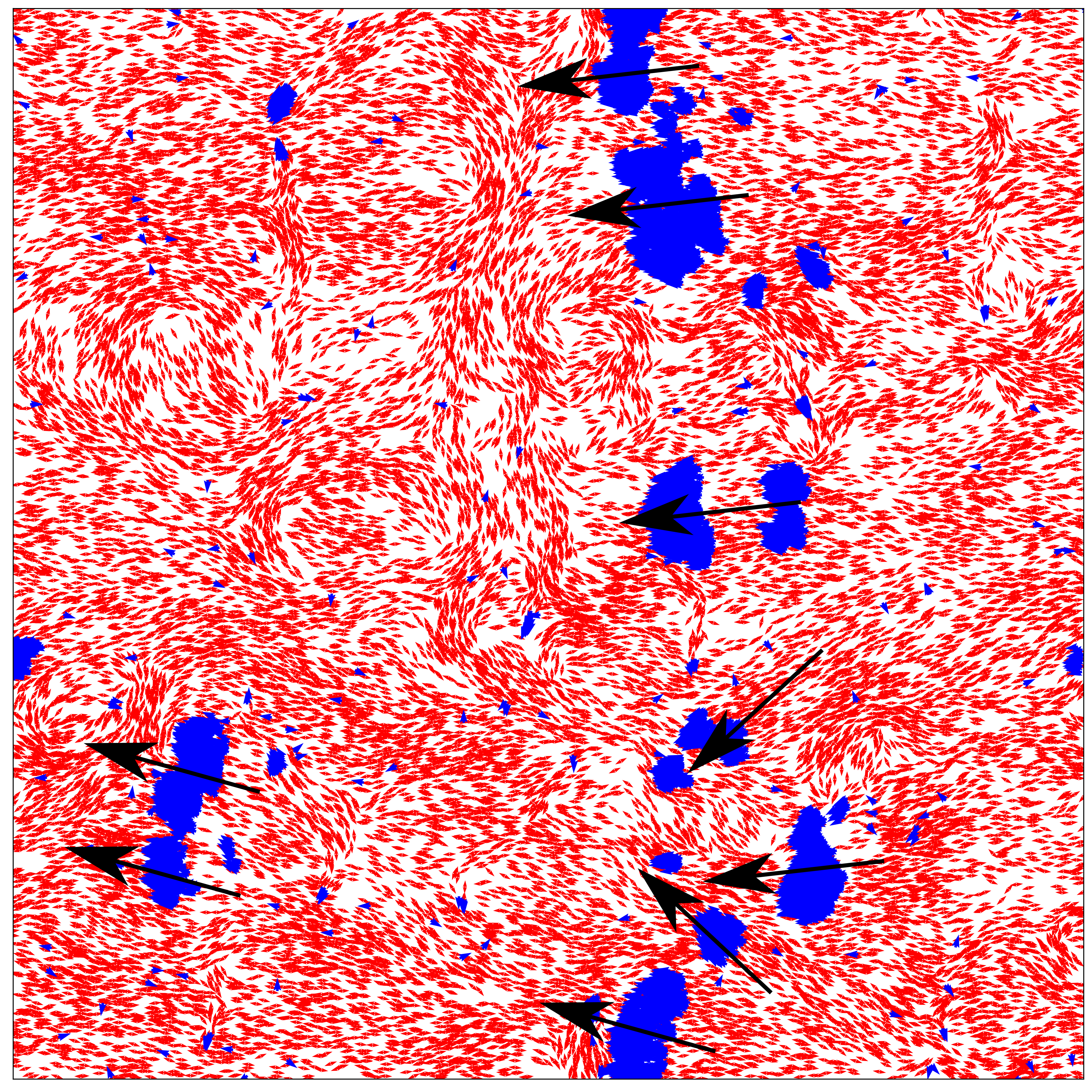}}&{\includegraphics[width=6cm]{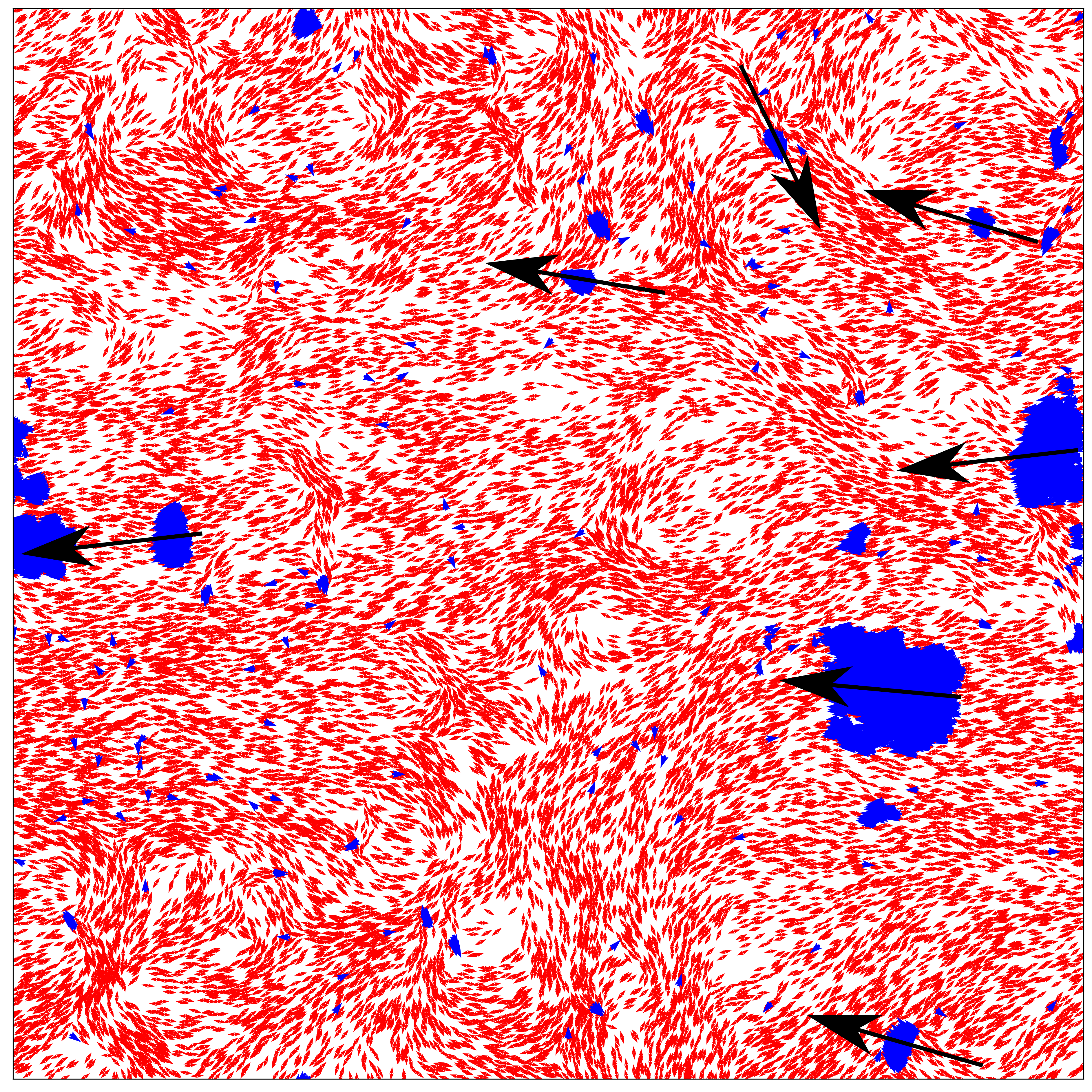}}&{\includegraphics[width=6cm]{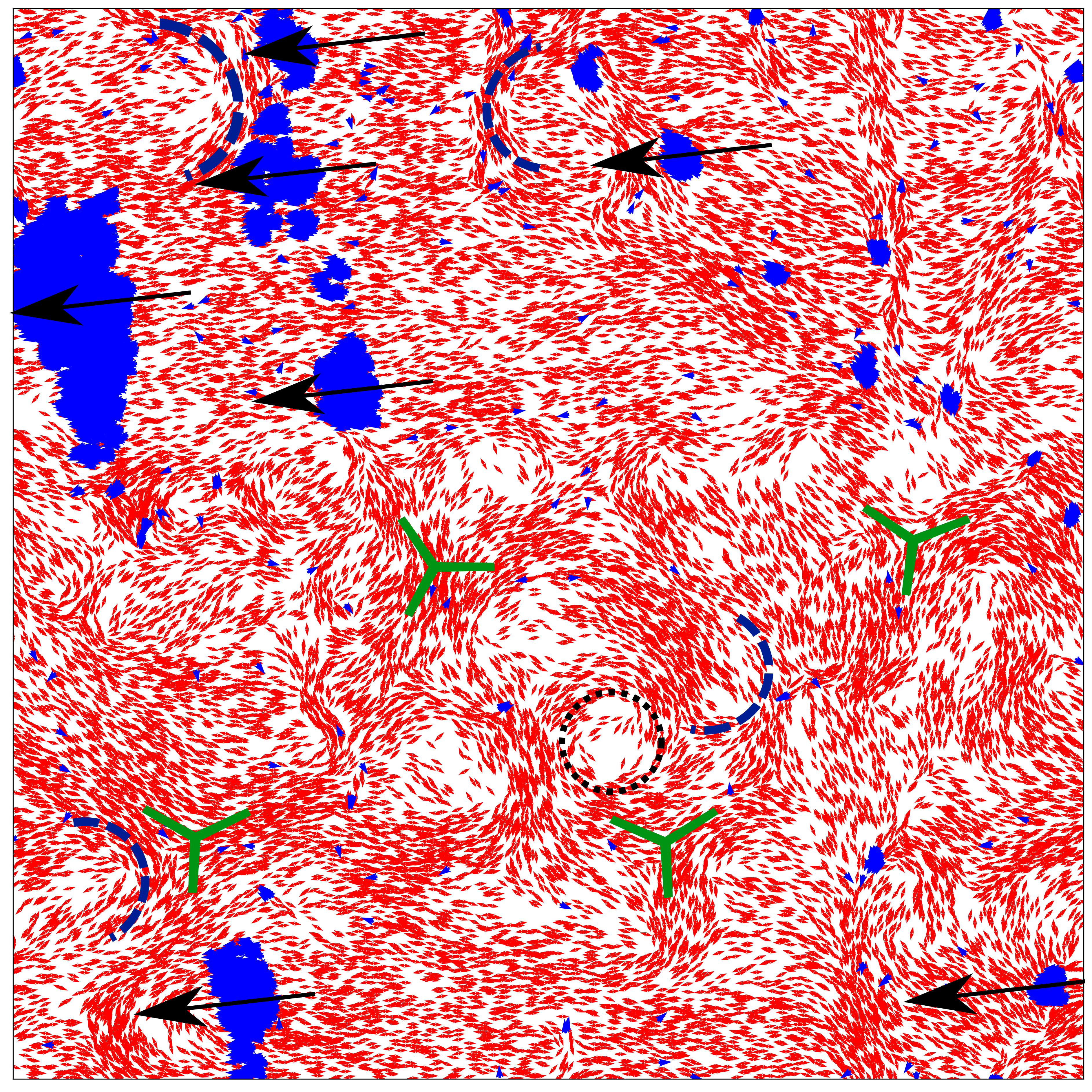}}
\end{tabular}
\caption{(Left-right) Snapshots of the mixture in the third phase, taken at time intervals of $5000$ time steps in the steady state. It illustrates the mutual alignment and persistent motion of large clusters of polar swimmers. The colours of the particles are the same as in Fig \ref{snapshots}. The black arrows represent the direction of the cluster of polar swimmers. The last panel shows $+\frac{1}{2}$ defects (dashed lines), $-\frac{1}{2}$ defects (solid lines), and higher order structures (dotted lines).}
\label{thirdphase}
\end{figure*}

In the first phase of the mixture, the polar swimmers have a low polar order parameter, with a sharp exponentially decaying velocity auto-correlation  which decays
to $0$ with a very short tail as shown in Fig. \ref{vcf}(a-b), and diffusive motion with $\Delta_p(\rho_p,t) \sim t$ and $\beta(t) \sim 1$ (Fig. \ref{polarmsd}). This behaviour is consistent with that of a random walker in a two-dimensional space, and is confirmed through real-space snapshots and the distribution of $\theta_p$ in the steady state (Fig. \ref{snapshots}(a)), which shows a uniform distribution of random spikes in $N(\theta_p)$.\par

In the second phase of the mixture, the polar swimmers show a moderate value of $V(\rho_p)$. The VACF of the polar swimmers decays exponentially to $0$, but with a much longer tail than the swimmers in the first phase. The correlation time $\tau(\rho_p)$ increases sharply, as seen in Fig. \ref{vcf}(b). Given the properties of $C(\rho_p,t)$, we fit it to the form $C(\rho_p,t)=(1-C_0(\rho_p))e^{\frac{-t}{\tau(\rho_p)}}+C_0(\rho_p)$. The exponent $\beta(t)$ of $\Delta_p(\rho_p, t)$ also shows that the motion of the swimmers is super-diffusive $\beta(t) > 1$ as seen in Fig. \ref{polarmsd}. This shows that the motion of the polar swimmers in the second phase is persistent for a short time-scale, but not over long periods. Real-space snapshots at $\rho_p=0.05$ show that the polar swimmers in this phase form multiple clusters which are not mutually aligned (Fig. \ref{snapshots}(b)).\par

The polar swimmers in the third phase have a high value of $V(\rho_p) \sim 1$, approaching a perfectly ordered state. They also show an exponential decay in $C(\rho_p, t)$, but the function decays to a non-zero constant value $C_0(\rho_p)$, rather than $0$, as seen in the earlier two phases Fig. \ref{vcf}. The correlation time $\tau(\rho_p)$ approaches a finite value in this phase. The motion of the polar swimmers in this phase can also be described as ballistic, with the exponent $\beta(t) \simeq 2$ Fig. \ref{polarmsd}. This suggests that the bulk of polar swimmers in the third phase undergo a highly persistent ballistic motion, with no significant deviation due to the presence of apolar rods. These dynamics are consistent with the highly ordered polar flocks described in \cite{vicsekmodel}. Snapshots in Fig. \ref{thirdphase} show that the polar swimmers form large, mutually aligned clusters which are persistent in their motion, and move along the direction of nematic alignment. The variation of $\tau(\rho_p)$ and $C_0(\rho_p)$ with $\rho_p$ is shown in Fig. \ref{vcf}.

\subsection{Structural Defects in Active Nematics}

The first phase of the mixture largely shows the same properties as a system comprising  of pure  nematic rods. Given the noise strength, strongly ordered dense bands of apolar rods are observed which are parallel to the direction of bulk apolar alignment.\par

\begin{figure*}
\centering
\begin{tabular}{cccc}
\fbox{\includegraphics[width=4.25cm]{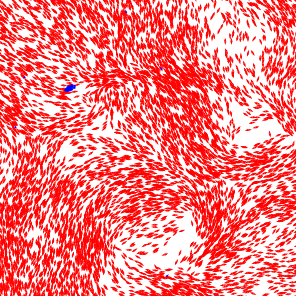}}& \fbox{\includegraphics[width=4.25cm]{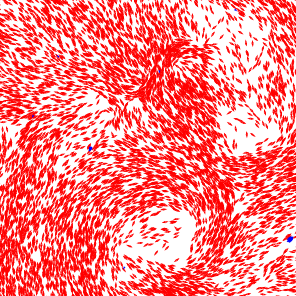}}& \fbox{\includegraphics[width=4.25cm]{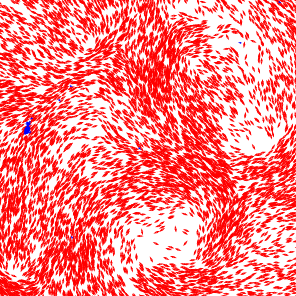}}& \fbox{\includegraphics[width=4.25cm]{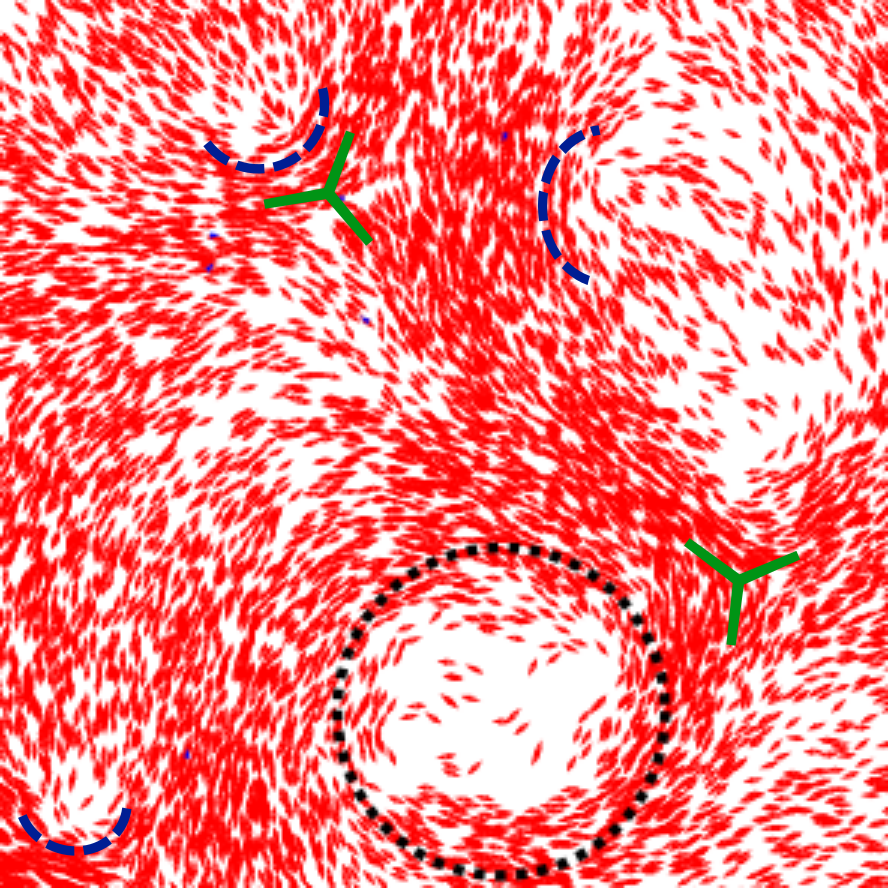}}
\end{tabular}
\caption{(Left-right) Zoomed in snapshots of the mixture in the second phase $(\rho_p=0.005)$, taken at intervals of 500 time-steps in the steady state. The snapshots show one quarter of the whole system and the colours of the particles are the same as in Fig. \ref{snapshots}. The last panel shows $+\frac{1}{2}$ defects (dashed lines), $-\frac{1}{2}$ defects (solid lines), and higher order structures (dotted lines).}
\label{phase2defects}
\end{figure*}

As $\rho_p$ is increased and approaches the critical range for the transition from the first phase to the second, it is observed that the polar swimmers begin clustering. The clusters of polar swimmers formed in the second phase are not mutually aligned and can strongly influence the orientation of apolar rods in regions where the density of apolar rods is lower than the density of the polar cluster. This leads to defect formation in nematic ordering, which in turn leads to high-density regions of apolar rods. Once formed, the dense regions of apolar rods bend the motion of the polar clusters that pass them. If the size of the polar cluster is small, they get rotated in the direction of the defects. However, if a cluster of polar swimmers is large enough, then it can further distort the defect structure. This dynamics keeps reoccurring in the system. Hence, the steady state of the system shows the formation, destruction and reformation of $\pm\frac{1}{2}$ topological defects in nematic ordering and higher order (circular) structures in the form of dense rings of apolar rods. The core of the circular structure is void of particles and the density of rods is highest at the circumference of the structure. As we get deeper into the second phase, multiple uncorrelated clusters of polar swimmers are formed. Two or more of such clusters cumulatively cause defects in nematic ordering, whose size is much larger than the size of the clusters themselves. A time series of snapshots of the mixture in the second phase ($\rho_p=0.005$) is shown in Fig. \ref{phase2defects}. The last panel of Fig. \ref{phase2defects} shows the presence of $\pm \frac{1}{2}$ and higher order defects in the second phase. $-\frac{1}{2}$ defects and the circular structures are more clearly visible in steady state of the system.\par
As $\rho_p$ is further increased, the polar swimmers form large clusters which are mutually aligned in a polar manner and show a high VACF. These large persistent clusters sweep the lattice repeatedly, and they have a comb-like effect on the apolar rods. They tend to enforce a global field for nematic alignment which is parallel to their own motion. Smaller stray clusters of polar swimmers still exist, and they prevent perfect nematic ordering of apolar rods. A time-series of real space snaphots of  the mixture in the third phase is shown in Fig. \ref{thirdphase}, which shows the time evolution of the defects, which are much smaller in size in comparison to the second phase. The  $\pm\frac{1}{2}$ and higher order defects are marked in the last panel of Fig. \ref{thirdphase}. $-\frac{1}{2}$ defects and the circular structures are more clearly visible in steady state of the system.  The circular structures in this phase are much smaller in size than the ones found in the second phase of the mixture.

\subsection{Number Fluctuations}

One of the characteristics of pure active nematics in the ordered state is the presence of large density fluctuations, also called as giant number fluctuations (GNF) \cite{chateminimal, shradhaprl2007,  sdeyprl2012}. The density fluctuation is measured by calculating the number fluctuation $\Delta N$ in boxes of different sizes. The GNF as found for the previous studies of pure systems is also observed in the first phase of the mixture. The fluctuations are partially supressed in the second phase of the mixture, which does not show large bands. However the large scale defects which lead to dense arrangements of apolar rods are still a cause for fluctuations. The third phase shows the smallest fluctuations, and even within the third phase fluctuations are further supressed with an increase in $\rho_p$.
\begin{figure}[H]
	\includegraphics[width=8cm]{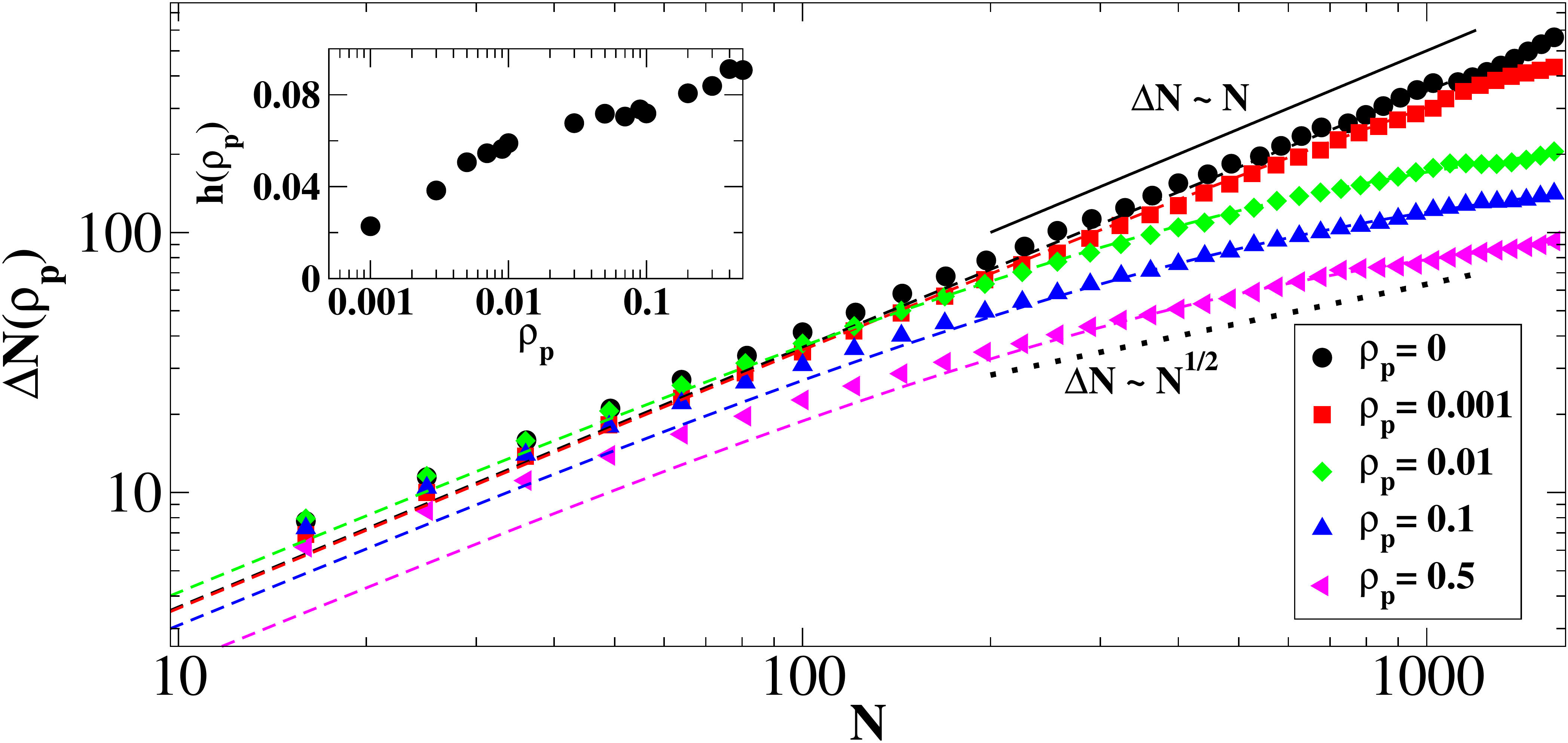}
	\caption{Variation of number fluctuation $\Delta N$ with $N$ for $\eta=0.2$ and $L=160$. The dashed line represents the fitted function as given in Eq. \ref{eqgnf}. (Inset) Variation of $h$ with $\rho_p$}
\label{numberfluc}
\end{figure}
The effect of ordered polar swimmers cluster can be assumed as an ordered field and result of number fluctuation is compared with the previous linearised calculation of $\Delta N$ for active nematic with an external field \cite{shradhaphiltrans2014}.\par

\begin{equation}
\frac{\Delta N}{N}=\frac{N}{(a+Nh^{2})^2}
\label{eqgnf}
\end{equation}

In the above expression Eq. \ref{eqgnf}, $a$ is a system dependent constant, which in the pure active nematic system depends on the diffusivity of the system and $h$ is the strength of the external field. In Fig. \ref{numberfluc} we plot the number fluctuation $\Delta N$ vs. $N$ for densities $\rho_p=0,0.001,0.01,0.1,0.5$, and show the comparison to the analytical expression derived for the active nematic with external field. For all densities the data fits well with the analytical expression. From the fitted expression in Eq. \ref{eqgnf} we find that the value of $a$ does not show significant variation on increasing $\rho_p$ (data not shown) and $h$ increases with $\rho_p$ as shown in Fig. \ref{numberfluc} (inset). The increase in $h$ with $\rho_p$ supports the hypothesis that the large clusters formed in the third phase act as a sweeping orienting field for the active nematics.
\section{Discussion}
\label{section:discussion}

To conclude, we have studied a mixture of self-propelled apolar rods and polar swimmers on a two-dimensional substrate.  The density of apolar rods is kept fixed and the density of polar swimmers is tuned from very small to moderate values such that polar swimmers are in minority. We find three distinct phases: The first phase is observed when the density of polar swimmers in the system is very low $\rho_p \sim 0.001$, and the behaviour of the apolar rods in this phase is largely similar to a pure system of apolar rods. However, the behaviour of the system changes significantly when the density of polar swimmers is large enough for polar swimmers to cluster, and the system transitions to a second phase. In the second phase of the mixture, clusters of polar swimmers are not aligned with each other, and subsequently destroy the nematic ordering of apolar rods. The transition from the first phase to the second phase is a first order transition with the coexistence of ordered and disordered realisations of apolar rods for the same system parameters near the critical $\rho_p$. Further increasing the density of polar swimmers, it is observed that beyond a threshold density of polar swimmers, the system transitions to a third phase where clusters of polar swimmers are able to exhibit high polar order. The large, mutually aligned clusters of polar swimmers formed in this phase are found to contribute to nematic ordering, and suppress the giant number fluctuations seen in a pure system of apolar rods and the first phase of the mixture. However, stray clusters of polar swimmers in this phase still lead to a markedly lower level of nematic ordering in the apolar rods. The transition from the second to the third phase is a continuous transition, and the system transitions from a disordered state to a short ranged ordered state.\par

Interestingly in the second phase we find formation of large topological defects and higher order structures in nematic ordering. The steady state is characterised by the formation and destruction of these defects and structures. The presence of clusters of polar swimmers are responsible for the formation and reformation of such structures. Such a dynamical steady state is absent when self-propelled apolar rods are replaced by passive apolar rods or the backgroud apolar rods are in equilibrium, as seen in LLC systems \cite{zhoullc1, zhoullc2}. Hence our predictions can be used for the early detection of polar swimmers in self-propelled apolar mixtures. It can also be used for temporary trapping of swimmers at the core of the defects.\par

While the highest density of polar swimmers investigated in this study is smaller than the density of apolar rods, preliminary investigation of higher densities of polar swimmers suggest that stray clusters of polar swimmers remain even when the population of both species is equal. This will have to be confirmed through a more rigorous study. It presents an interesting problem, because it is expected that a system where the density of polar swimmers is much higher than the density of apolar rods should show similar traits to a pure system of polar swimmers, which might lead to another phase transition at that extremum.\par

\section{Acknowledgement}
SM and PS would like to thank Tamás Vicsek, Robert Pelcovits and Rakesh Das for giving useful comments on the manuscript. The support and the resources provided by PARAM Shivay Facility under the National Supercomputing Mission, Government of India at the Indian Institute of Technology, Varanasi are gratefully acknowledged by SM and PS.  SM  thanks  DST-SERB  India,  ECR/2017/000659 for financial support. SM and PS also thank the Centre for Computing and Information Services at IIT (BHU), Varanasi.
\bibliographystyle{ieeetr}
\bibliography{bibliography}
\end{document}